\documentclass[twocolumn]{aastex63}
\usepackage{sidecap}
\usepackage{wrapfig}
\usepackage{float}
\usepackage{xcolor}
\usepackage{lineno}
\usepackage{soul}

\accepted{Oct 19, 2021}
\submitjournal{ApJ Letters}

\shorttitle{KELT-9 Dayside}
\shortauthors{Kasper et al.}
\graphicspath{{./}}

\begin{document}
\title{Confirmation of Iron Emission Lines and Non-detection of TiO on the Dayside of KELT-9b with MAROON-X}
\correspondingauthor{David Kasper}
\email{kasperd@uchicago.edu}

\author[0000-0003-0534-6388]{David Kasper}
\affil{Department of Astronomy \& Astrophysics, University of Chicago, 5640 South Ellis Avenue, Chicago, IL 60637, USA}

\author[0000-0003-4733-6532]{Jacob L.\ Bean}
\affil{Department of Astronomy \& Astrophysics, University of Chicago, 5640 South Ellis Avenue, Chicago, IL 60637, USA}

\author{Michael R.\ Line}
\affil{School of Earth and Space Exploration, Arizona State University, Tempe, AZ 85281, USA}

\author[0000-0003-4526-3747]{Andreas Seifahrt}
\affil{Department of Astronomy \& Astrophysics, University of Chicago, 5640 South Ellis Avenue, Chicago, IL 60637, USA}

\author[0000-0002-4410-4712]{Julian St\"urmer}
\affil{Landessternwarte, Zentrum f\"ur Astronomie der Universit\"at Heidelberg, K\"onigstuhl 12, 69117 Heidelberg, Germany}

\author[0000-0002-1321-8856]{Lorenzo Pino}
\affil{INAF-Osservatorio Astrofisico di Arcetri Largo Enrico Fermi 5
I-50125 Firenze, Italy}

\author{Jean-Michel D\'esert}
\affil{Anton Pannekoek Institute for Astronomy, University of Amsterdam, 1098 XH Amsterdam, The Netherlands}

\author[0000-0002-7704-0153]{Matteo Brogi}
\affil{Department of Physics, University of Warwick, Coventry CV4 7AL, UK}
\affil{INAF-Osservatorio Astrofisico di Torino, Via Osservatorio 20,
I-10025 Pino Torinese, Italy}
\affil{Centre for Exoplanets and Habitability, University of Warwick,
Coventry, CV4 7AL, UK}


\begin{abstract}
We present dayside thermal emission observations of the hottest exoplanet KELT-9b using the new MAROON-X spectrograph. We detect atomic lines in emission with a signal-to-noise ratio of 10 using cross-correlation with binary masks. The detection of emission lines confirms the presence of a thermal inversion in KELT-9b's atmosphere. We also use M-dwarf stellar masks to search for TiO, which has recently been invoked to explain the unusual \textit{HST}/WFC3 spectrum of the planet. We find that the KELT-9b atmosphere is inconsistent with the M-dwarf masks. Furthermore, we use an atmospheric retrieval approach to place an upper limit on the TiO volume mixing ratio of 10$^{-8.5}$ (at 99\% confidence). This upper limit is inconsistent with the models used to match the WFC3 data, which require at least an order of magnitude more TiO, thus suggesting the need for an alternate explanation of the space-based data. Our retrieval results also strongly prefer an inverted temperature profile and atomic/ion abundances largely consistent with the expectations for a solar composition gas in thermochemcial equilibrium. The exception is the retrieved abundance of Fe$^+$, which is about 1-2 orders of magnitude greater than predictions. These results highlight the growing power of high-resolution spectrographs on large ground-based telescopes to characterize exoplanet atmospheres when used in combination with new retrieval techniques.
\end{abstract}



\section{Introduction} \label{sec:intro}
Since its discovery just a few years ago, the ultra-hot Jupiter KELT-9b has become a touchstone for studying the extreme physics and chemistry in the atmospheres of highly-irradiated planets \citep{Gaudi2017}. The key observational results to date for the atmosphere of this planet include the detection of a wide range of neutral and ionized atomic species \citep{hoeijmakers18, hoeijmakers19, cauley19, yan19, turner20,pino20}, an extended envelope of excited hydrogen \citep{yan18}, dynamics and variability \citep{cauley19}, and heat transport from H$_{2}$ dissociation and recombination \citep{mansfield20,wong20}. The overall picture that is painted by these observations is consistent with the predictions of models that assume roughly solar composition, thermochemical equilibrium, and standard atmospheric physics \citep{kitzmann18, lothringer18, bell18, parmentier18, tan19, fossati21}. This broad brushstroke agreement is a major success of exoplanet atmosphere theory, and it enables using the data and models together to explore the detailed properties of KELT-9b's atmosphere.


Despite most observations of KELT-9b agreeing with established atmospheric models, recent measurements of the planet's dayside thermal emission defy expectations. \citet{Changeat2021} used \textit{Hubble Space Telescope} WFC3 data to show that the planet's low-resolution thermal emission spectrum in the near-infrared is highly featured. They suggested that these features could be explained by opacity from the molecules TiO, VO, and FeH. The presence of molecules on the dayside of the planet is surprising because the infrared photospheric temperature has been measured to be $\sim$4,500\,K \citep{mansfield20,wong20} and molecules should to be dissociated at these high temperatures \citep{kitzmann18, lothringer18}. \citet{Changeat2021} point out that their derived molecular abundances are many orders of magnitude greater than what is predicted by thermochemical equilibrium. Therefore, these results would likely require a high metallicity composition and/or non-equilibrium chemistry, which would be novel results for ultra-hot Jupiter atmospheres.


We present ground-based high-resolution observations of the dayside emission from KELT-9b that were obtained using the new MAROON-X instrument on the Gemini North telescope \citep{seifahrt16, seifahrt18, seifahrt20}. We analyzed these data to test the claimed detection of molecules by \citet{Changeat2021} and to further explore the chemistry of this unique planet. MAROON-X's red optical bandpass was designed to cover the rich molecular spectra of M dwarf stars for extreme precision radial velocity measurements. The spectra of M dwarfs in this wavelength range are dominated by the same molecules that are suggested by the WFC3 data, so the instrument is ideally suited for this investigation.

We present our observations in \S\ref{sec:obs}. We describe two analyses of the data in \S\ref{sec:template} and \ref{sec:retreival} that yield the detection of the planet's atmosphere and constraints on its properties. We conclude with a discussion of the results in \S\ref{sec:discussion}.

\section{Observations} \label{sec:obs}
We used MAROON-X to obtain 5.9\,hours of data (7.7\,hours total observing time including overheads) during pre-secondary eclipse phases of KELT-9b over three nights in May and June 2020 with both the `blue' (500 -- 663\,nm) and `red' (654 -- 920\,nm) channels. This was the first regular observing run of the instrument following commissioning and science verification in 2019. The two MAROON-X detectors read out at different rates, so we used different exposure times to maintain a constant cadence throughout the observations (blue channel: 200\,s, red channel: 180\,s). We also continued observing past the start of the secondary eclipse during two of the nights to obtain data when the planet wasn't visible as a control (1.1\,hours open-shutter time, 1.25\,hours total observing time including overhead). Table~\ref{tab:obs_log} gives a log of the observations.


\begin{deluxetable*}{lccccccccclcc}
\tabletypesize{\scriptsize}
\tablecolumns{13}
\tablewidth{0pc}
\tablecaption{\label{tab:obs_log} Log of observations}
\tablehead{
 \colhead{UT Date}  & \colhead{Exposures} &  \colhead{Planet Phase Range} & \colhead{Airmass} & \colhead{Conditions} & \colhead{Seeing} & \colhead{Average Blue Arm SNR}
}
\startdata
2020 May 28 12:15 $\rightarrow$ 14:55 & 37 & 0.383 $\rightarrow$ 0.458 & 1.21 $\rightarrow$  1.07 & some passing clouds & 0.5 - 0.8\arcsec & 140\\ 
2020 May 31 10:15 $\rightarrow$ 14:07 & 53 & 0.353 $\rightarrow$ 0.461 & 1.70 $\rightarrow$ 1.10 & clear & 0.3 - 0.5\arcsec & 200\\
2020 June 03 10:18 $\rightarrow$ 12:53 & 36 & 0.379 $\rightarrow$ 0.444 & 1.60 $\rightarrow$ 1.12 & light clouds early & 0.4 - 0.6\arcsec & 191\\
\enddata
\end{deluxetable*}		

\section{Detection of the Planet's Atmosphere} \label{sec:template}
We performed two analyses of our data to characterize KELT-9b's atmosphere. The first analysis used a cross-correlation function (CCF) technique with line-weighted binary templates. We broadly followed the \citet{pino20} Appendix A \& B approach to create a time (i.e. phase) evolving spectrum containing planet lines normalized to the planet plus star continuum.

We used nine stellar templates as masks for the CCF analysis. These masks correspond to F9, G2, K2, K6, M0, M2, M3, M4, and M5 stellar types and come from the ESPRESSO data reduction pipeline (``ESPRESSO DRS''\footnote{\url{http://eso.org/sci/software/pipelines/}}). The masks span the transition from ionized and neutral atoms to molecules as the dominant species in hydrogen-dominated atmospheres and are typically considered as plausible proxies for hot Jupiters. These masks were built from observed stellar spectra \citep[e.g.,][]{suarez20}. Therefore, they shouldn't suffer from errors in line positions and strengths that sometimes hinder the use of theoretical templates, which has been a particular problem for searches for TiO \citep{hoeijmakers15, nugroho17, herman20, serindag21}. The M dwarf masks also contain lines from VO, which is another molecule that was proposed by \citet{Changeat2021}.

For this analysis, we calculate the detection signal-to-noise ratio (SNR) in our CCF maps by normalizing the peak value with the mean and standard deviation of the values in the rest of the map after applying an iterative 3-sigma clipping. We confirm that this approach gives similar (but slightly higher) detection SNR as the Welch's T-test that was developed for CCF analyses by \citet{Brogi2013}.

\begin{figure}[t!]
\begin{center}
\includegraphics[width=\linewidth]{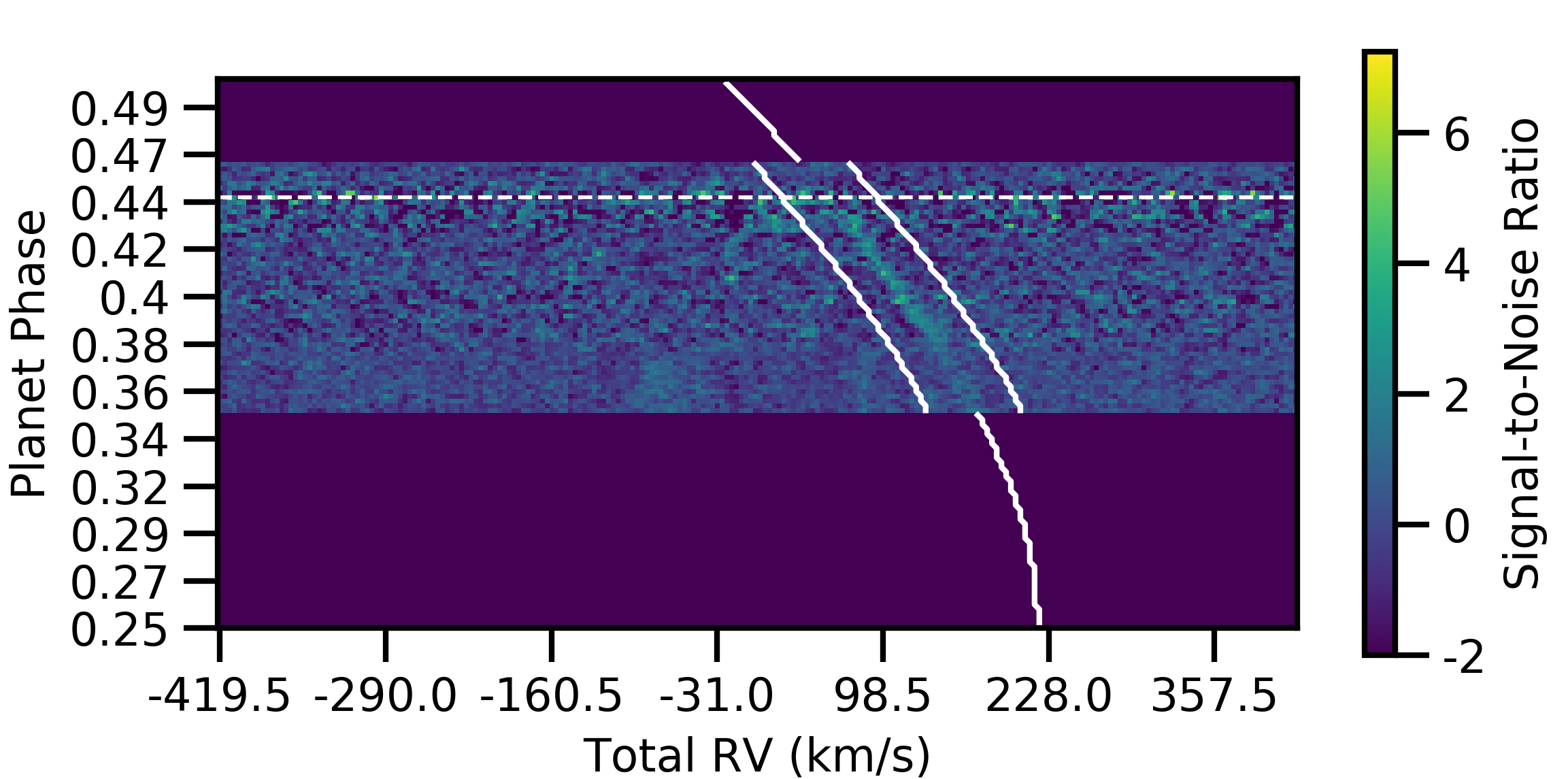}
\caption{\textbf{SNR of the planetary atmospheric signal with the G2 stellar template.} This data set is the co-addition of the three nights of observations using the MAROON-X blue channel. The data series were co-added by nearest phase matching to the May 31 night as that dataset contained the complete phase ranges of the other two nights. This procedure gave minimal signal smearing caused by a maximum phase mismatch of 0.1\% (i.e. 130\,sec). The color quantifies the number of standard deviations from the mean response within the phases observed (with additional, non-observed phases shown as zero SNR for clarity). The horizontal white dashed line indicates the beginning (i.e., first contact) of the secondary eclipse as calculated from the most recent ephemeris \citep[phase 0.4449;][]{wong20,mansfield20}. The white solid lines indicate the predicted trace of the planetary atmosphere in planet+system velocity with a $\pm 10$\,km\,s$^{-1}$ offset applied during the observed phases.}
\label{fig:ccf_coadd}
\end{center}
\end{figure}

As shown in Figure~\ref{fig:ccf_coadd}, the CCF exposure matrix of the combined data series demonstrates that the planet's atmosphere is detected at high confidence along the expected radial velocity trace of the planet. The signal disappears at secondary eclipse when the planet is blocked by the star. The relatively weaker signal strength along the trace at the earliest observed phases is due to these phases only being covered by one of our three data sets. Nevertheless, even here the strength of the detection is high \citep[compare to Figure 2 in][]{pino20}. As previously discovered by \citet{pino20}, we see a positive correlation between the masks and the data, which indicates that the spectral lines are in emission due to a thermal inversion in the planet's dayside atmosphere.

\begin{figure}[t!]
\begin{center}
\includegraphics[width=\linewidth]{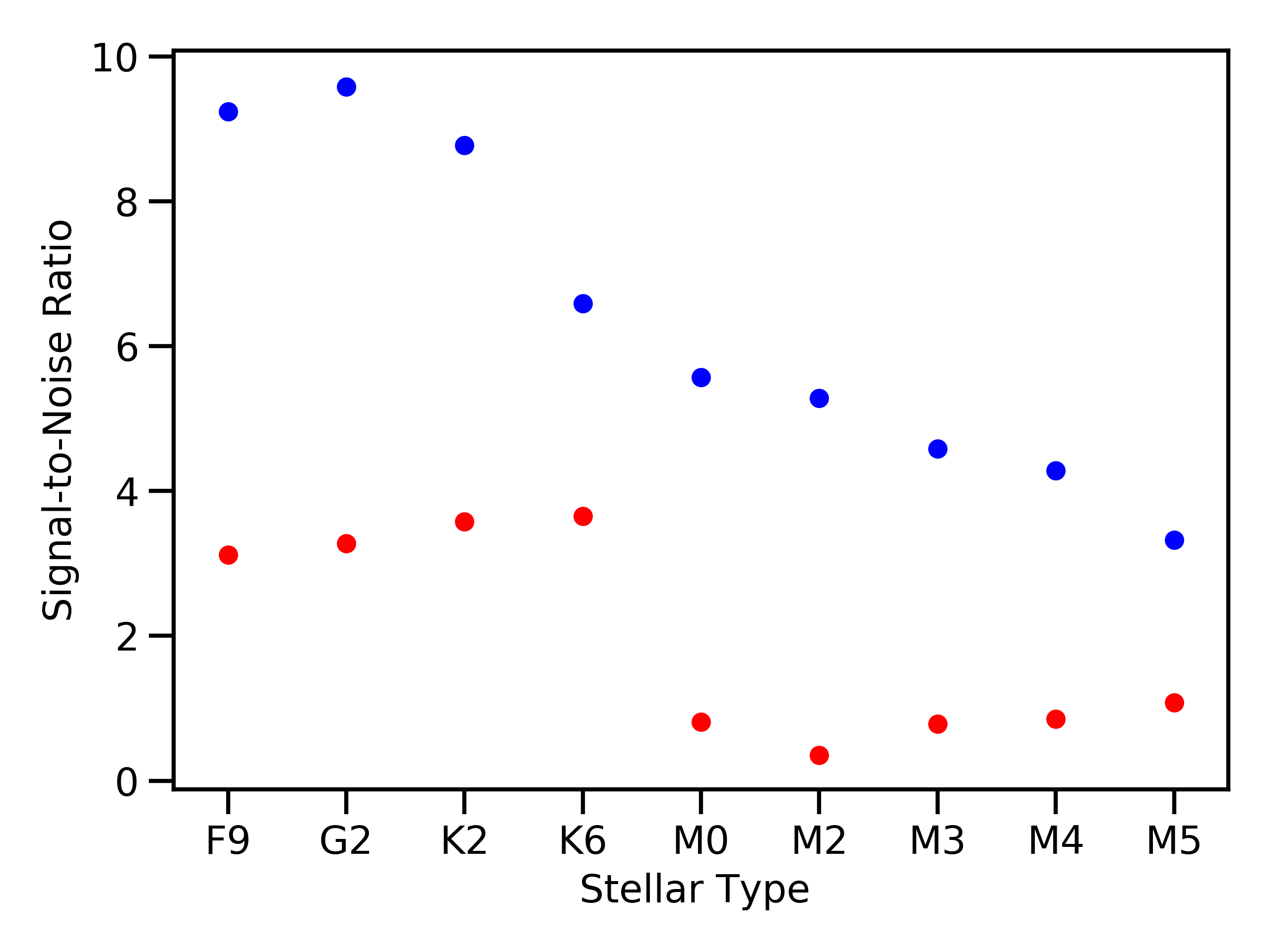}
\caption{\textbf{Detection SNR of the planetary atmospheric signal as a function of template stellar type.} The peak SNR for the co-addition of the three observation sequences are shown for each arm independently. The absolute peak in SNR (at $9.6\,\sigma$) occurs in the blue channel (blue points), utilizing the G2 mask. The blue channel broadly has an inverse relationship between signal strength and later spectral type. The red channel (red points) has a marginally significant response at early spectral types and is indistinguishable from noise at later spectral types.}
\label{fig:ccf_comp}
\end{center}
\end{figure}

Figure \ref{fig:ccf_comp} shows the detection SNR as a function of mask spectral type for the blue and red channel data. All of the high SNR detections were found near the expected planet and system velocities, which points to the nature of the recovered signal as planetary in origin. This plot demonstrates the relative importance of the signal in the blue channel over the red channel in the confidence of detection despite very similar data quality (photon-limited SNR within 2\%) in the two channels. The earlier spectral type masks give detection confidences approaching $10\sigma$ in the blue channel. The M-type masks give substantially lower detection confidences in the blue channel, and do not yield a significant detection in the red channel. These results suggest that KELT-9b's dayside spectrum is dominated by the atomic lines (primarily neutral iron, see \S\ref{sec:retreival}) that are more prevalent in the earlier-type masks and that are concentrated in the bluer part of the MAROON-X bandpass. The lack of a significant detection with the M-type masks in the red channel disfavors the presence of molecules like TiO.


\begin{figure*}
\begin{center}
\includegraphics[width=\linewidth]{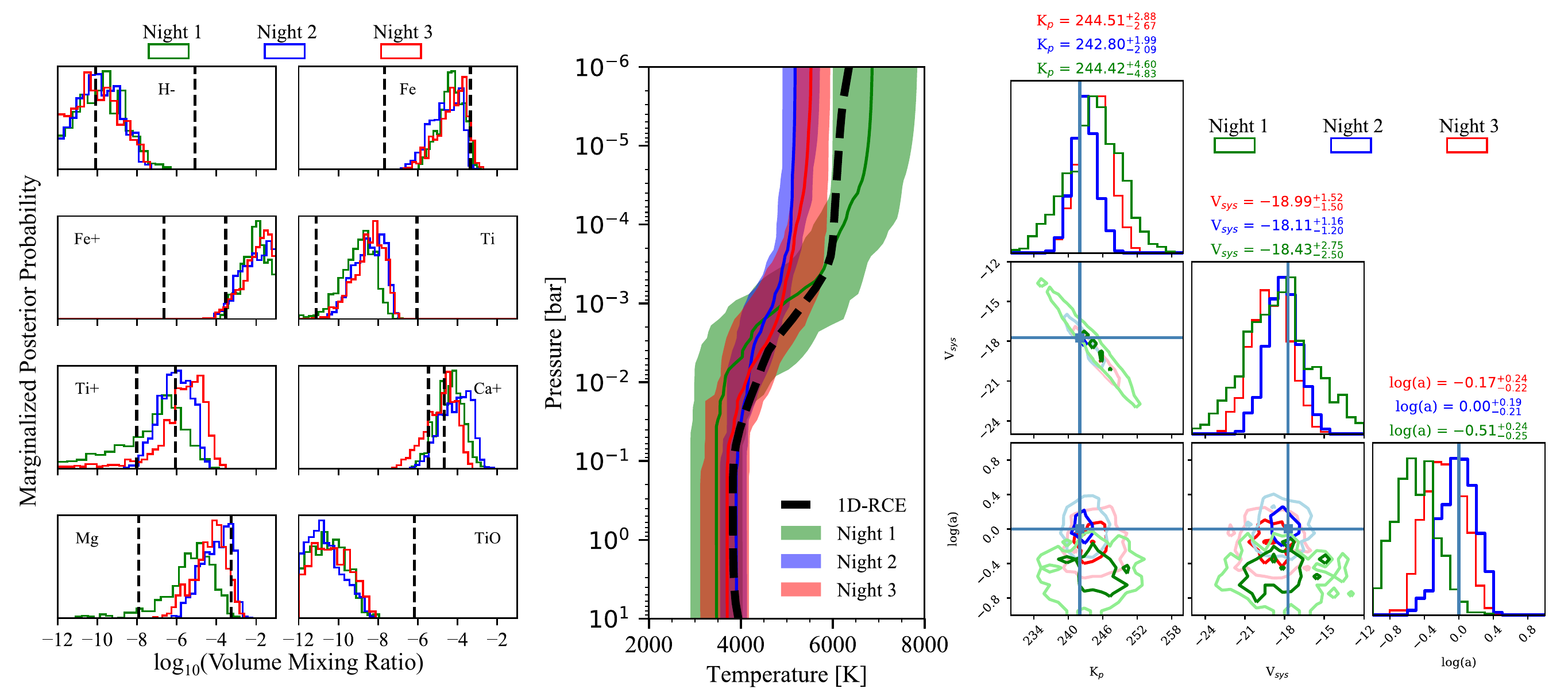}
\caption{\textbf{Retrieval results for KELT-9b}. The left panels show the volume mixing ratios for key chemical species. The vertical dashed lines represent the range of abundances encountered in chemical equilibrium along the temperature profile for a 10$\times$ solar metallicity atmosphere \citep{pino20}. The central panel shows the derived temperature-pressure profiles (colored lines and shaded 1$\sigma$ confidence regions) along with the prediction from a 1D radiative-convective equilibrium model (black dashed line). The right panels show the derived planet and system velocities (solid lines are predicted values from previous constraints) along with the logarithm of the dilution factor, which is expected to be centered on zero.}
\label{fig:retreival}
\end{center}
\end{figure*}


\section{Retrieval Analysis} \label{sec:retreival}
In our second analysis we applied the \citet{BL19} cross-correlation retrieval framework to derive abundances and the vertical temperature structure of KELT-9b's atmosphere. We applied the retrieval approach to the three nights of data independently. We performed a ``free'' retrieval in that we didn't enforce radiative-convective or thermochemical equilibrium. We closely followed the approach described in \citet{BL19} and \cite{line21}, except we use principle component analysis \citep[PCA,][]{dek13, Pelletier2021} telluric/stellar removal procedure instead of the polynomial fitting based airmass detrending method. The following procedure is predicated on the assumption that the total signal can be written as
\begin{equation}
S(\lambda,t)=\left(1+\frac{F_p(\lambda,t)}{F_{\star}(\lambda,t)}\right)\mathcal{T}(\lambda,t),    
\end{equation}
where $F_p/F_{\star}$ is the secondary eclipse depth and $\mathcal{T}(\lambda,t)$ is a matrix that encapsulates the time and wavelength dependent instrumental throughput, telluric transmittance, and stellar spectrum \citep[see Appendix A of][]{pino20}. 

To briefly summarize the procedure for a given night of data: (1) a singular value decomposition is performed on the data cube for each order (frames/phases/time x pixel/wavelength) with the first $N$ eigenvalues (we chose four, but the results are insensitive to the exact value) nulled to remove the dominant (non-planetary) contributions to $\mathcal{T}(\lambda,t)$. The remaining eigenvalues/vectors are then used to reconstruct the residual matrix that contains the planetary signal \citep[see section 2 of][]{dek13}. (2) Independently, the top $N$ eigenvectors/values are used to reconstruct the $\mathcal{T}(\lambda,t)$ absent the true planetary signal. (3) A planetary and stellar model are used to produce an $F_p/F_{\star}$ spectrum. (4) The results of steps (2) and (3) are combined to create ``model-injected" data (after the appropriate Doppler shift is applied to the model at each phase given the planetary, $K_p$, and system+barycenter velocities, $V_{sys}+V_{bary}$), and (5) the PCA is applied to this model-injected data to produce a model-injected residual matrix (analogous to what is done to the data in step (1)). (6) The post-PCA residual matrices from (1) and (5) are then cross-correlated and log-likelihood computed, frame-by-frame (spectrum by spectrum) within each order, over all orders and then summed, resulting in a single log-likelihood value for a given model spectrum and velocity pair.  Steps (3)-(6) are then repeated within a Bayesian parameter estimation tool at each iteration \citep[{\tt PyMultiNest};][]{feroz2008}). See \citet{BL19} and \citet{Pelletier2021} for more details on this analysis/retrieval procedure. The retrieval analysis is performed on the blue and red arm data simultaneously and took five days to run for each data series.


The forward model used to generate the planetary spectrum utilizes the three-parameter \citet{Guillot2010, Line2013} temperature-pressure profile (TP) parameterization, uniform with altitude mixing ratios for H$^-$, Fe, Fe$^+$, Ti, Ti$^+$,  Ca$^+$, Mg, and TiO, the planet Keplerian ($K_p$) and system velocities (V$_{sys}$), and a dilution/scale factor, $a$, for a total of 14 free parameters. We used the most recent ``TOTO'' line list for TiO \citep{mckemmish19}, H- bound-free opacities from \citet{John1988}, and the ``Kurucz" atomic data available through the HELIOS-K package \citep{Grimm2015, Grimm2021}. All cross-sections (atomic/molecular) are generated with the HELIOS-K package assuming a 0.01\,cm$^{-1}$ resolution (R$\sim$5 million) and a 100\,cm$^{-1}$ wing cut-off, over a grid of temperatures (12 points from 2000 -- 4000\,K) and pressures (18 points from $\mu$bar -- 300\,bar).  Cross-sections are then down-sampled through interpolation to a constant R\,=\,500k.  The core thermal emission radiative transfer was originally presented in \citet{Line2013}, and more recently in \citet{Zaleskey19} and \citet{line21}

Figure~\ref{fig:retreival} shows the results of the retrieval analysis. The results from looking at the three nights of data separately agree very well, and the planet and system velocities are found to match previous measurements. The derived TP profile quantifies the thermal inversion indicated by the analysis in \S\ref{sec:template}. The confidence interval on the retrieved TP profile encompasses the prediction of a 1D radiative-convective equilibrium calculation (assuming a dayside irradiation temperature of 4600\,K and a 10$\times$ solar metallicity) using the methods described in \citet{Arcangeli2018}.


The derived abundances of H$^-$, Fe, Ti, Mg, and TiO are found to be in agreement with the expectations for a roughly solar composition gas in thermochemical equilibrium. The retrieval indicates a stringent upper limit on the abundance of TiO of 10$^{-8.5}$ at 99\% confidence, again consistent with no detection at the abundances proposed by \citet{Changeat2021}.

\begin{figure}
\begin{center}
\includegraphics[width=\linewidth]{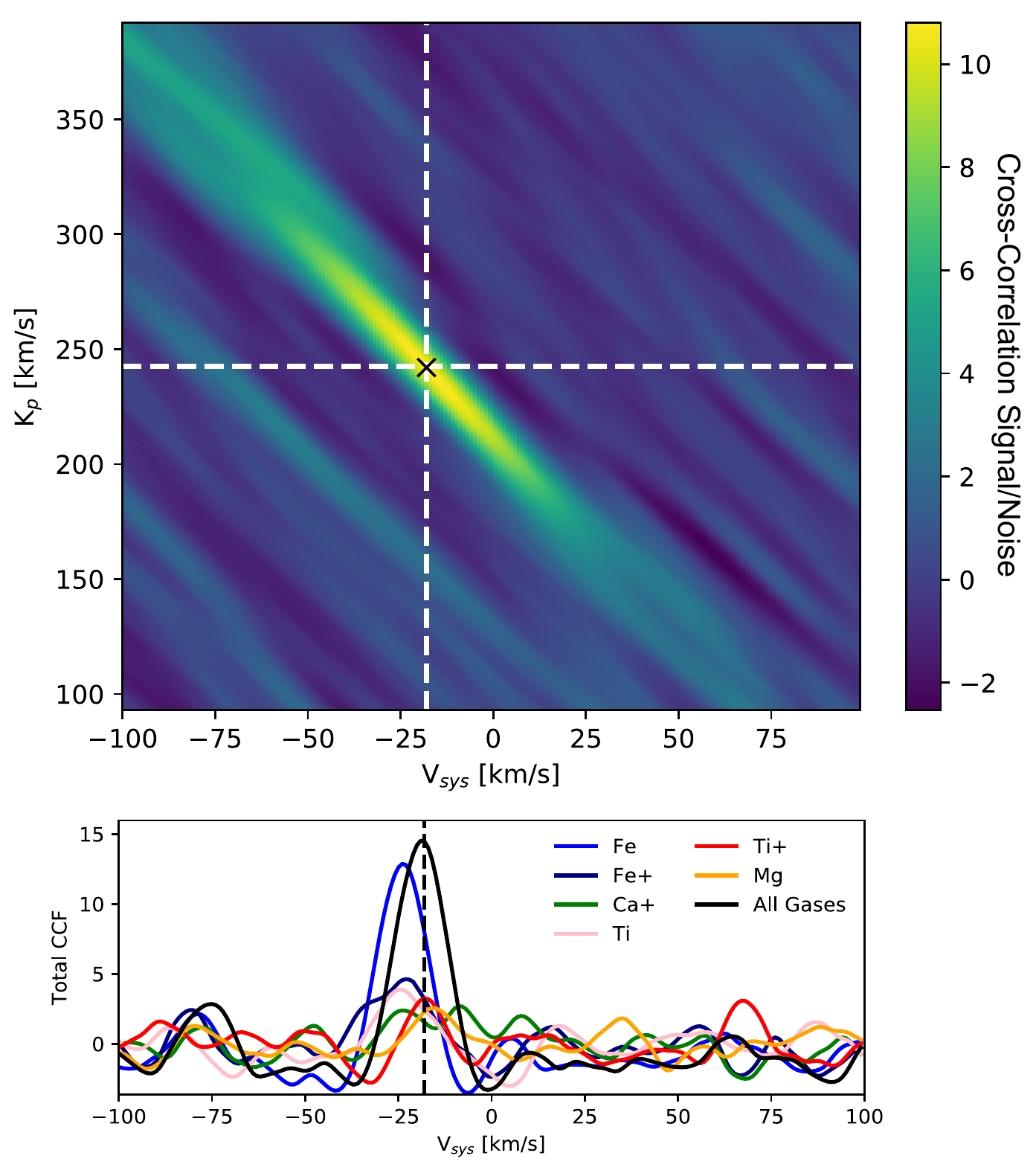}
\caption{\textbf{Cross-correlation SNR of the second night of data (May 31) with a randomly selected model drawn from the posterior as the template.} We cross-correlate the template model with each $K_p$-V$_{sys}$ pair to produce the top panel \citep[e.g.,][]{Birkby2018}. The SNR is calculated by normalizing the CCF map in the same way that was done for the binary mask calculation. The bottom panel is a cross-cut in V$_{sys}$ at the nominal $K_p$=242\,km\,s$^{-1}$ to illustrate the contributions from each individual species. This is done by cross-correlating the same template model with all gases removed but one (and the H$^{-}$ continuum).}
\label{fig:ccf_map}
\end{center}
\end{figure}

The retrieved abundances of Fe$^+$, and to a lesser extent Ti$^+$ and Ca$^+$, are higher than expected. Our model assumes local thermodynamic equilibrium (LTE) to compute both the electron energy level populations of the absorbing gasses (through the Boltzmann equation) and for predicting the partitioning of elements into different atoms, ions, and molecules for a given set of abundances (through the Saha equation). Photoionization, which is a non-LTE process, is known to impact the abundances of Fe and Mg species at low pressures in KELT-9b's atmosphere \citep{fossati21}. However, the total number of neutral and ionized atoms for a given element derived from a free retrieval would still equal the overall abundance of that element if it were just a question of changing the ionization balance. In this case the total abundance of neutral and ionized Fe would imply a roughly 100$\times$ solar metallicity, which we consider unlikely. Non-LTE may also impact the electron energy level populations, which would directly affect the radiative transfer calculations in our model, and thus is more difficult to account for.


Figure~\ref{fig:ccf_map} shows the second night's data cross-correlated (as a function of $K_p$ and V$_{sys}$ ) with the template spectrum randomly drawn from the retrieved posterior probability distribution. In the bottom panel we show the individual gas contributions along a slice in V$_{sys}$ at the nominal $K_p$. We note that this representation, which relies upon only a single template model, of the planetary signal (and individual gas contributions) does not fully reflect all of the information available in the data. While a gas may be constrained within the retrieval itself, it may not necessarily show up as a strong signal in the CCF representation. The retrieval analysis measures the change in the relative log-likelihood per change in gas abundance, which takes into account all of the corresponding degeneracies, whereas a simple CCF slice cannot fully represent for this.

From these figures we see that neutral iron is driving the detection. Interestingly, the CCF peak position differs by a few km\,s$^{-1}$ among the individual species. This is perhaps an indication that the species are distributed differently in the planet's atmospheres (i.e., they experience different effective velocities), thus contradicting our assumption of 1D geometry. The retrieval analysis maximizes the log-likelihood, which depends on the velocities and abundances of each gas. As such, the retrieval finds an average solution, but may not find the correct velocity pair for any individual gas itself. A similar, but larger, velocity offset was recently seen between Fe and TiO in WASP-33b \citep[which is a cooler planet than KELT-9b;][]{cont21}. If such an inhomogeneity exists in KELT-9b's atmosphere then that could provide another explanation for our surprising retrieval results for the ionized species. However, chemical inhomogeneities cannot explain the presence of TiO in the \textit{HST} WFC3 and the concomitant lack of a detection in our MAROON-X data because we don't see evidence of a TiO signal at any velocity consistent with a planetary origin.

\section{Discussion} \label{sec:discussion}
Our detection of atomic emission lines in the dayside spectrum of KELT-9b agrees with the previous results of \citet{pino20} that were obtained with HARPS-N. The larger aperture of the Gemini-N telescope yields higher signal-to-noise data and a stronger detection. The extended red optical coverage of MAROON-X permits a sensitive search for the molecules proposed by \citet{Changeat2021}, particularly TiO.  The TiO molecule has historically been challenging to detect in exoplanet atmospheres with high resolution techniques due to problems with the line list. To maximize our chance of a detection, we used the latest line list for TiO in our retrieval analysis \citep[see \S\ref{sec:retreival};][]{mckemmish19}, and we used templates derived from the observed spectra of M dwarfs (see \S\ref{sec:template}).

Our data strongly disfavor the high abundance of TiO that is needed to match the \textit{HST} WFC3 spectrum. We therefore see no evidence for molecules that could be due to non-equilibrium chemistry or a highly non-solar abundance pattern. Instead, our data reinforce and extend the previous ground-based high-resolution studies that highlight the presence of atomic rather than molecular species in KELT-9b's atmosphere.

The origin of the discrepancy between the ground-based high-resolution results and the WFC3 data remains a mystery. WFC3 results have been seen as the gold standard for transit spectroscopy measurements over the last decade. However, features in spectra that have come from single WFC3 transit/eclipse observations, like the KELT-9b dataset, have sometimes been over interpreted in our opinion. Our own analysis of the WFC3 data for KELT-9b yields a spectrum that is different than the \citet{Changeat2021} spectrum because we account for the stellar pulsations that were discovered by \citet{wong20} and \citet{mansfield20} (private communication from B. Jacobs). It remains to be seen whether this version of the WFC3 spectrum can be reconciled with the ground-based data. Ultimately, observations of KELT-9b's transmission, dayside emission, and phase curve with \textit{JWST} would be very valuable for furthering our understanding this extreme planet.

With a mass of $\sim$3\,M$_{\mathrm{Jup}}$, KELT-9b is not expected to have a particularly metal-enriched atmosphere if the mass-metallicity trend observed for the solar system giant planets is universal. Our retrieval results are largely consistent with this expectation. However, the derived abundance of Fe$^+$ is very high. We attribute this to the use of simplifying assumptions in the retrieval model (1D, constant abundances with altitude, LTE, etc.). Retrieval techniques for high-resolution studies of exoplanet atmospheres have advanced rapidly in the last few years \citep{bro17, pino18, BL19, shulyak19, gibson20, Pelletier2021}. The improvement in the data quality from new spectrographs on the largest ground-based telescopes, like MAROON-X on Gemini-N, motivates further improvement in retrieval techniques to reveal exoplanet atmospheres in greater detail.


\acknowledgments
The development of the MAROON-X spectrograph was funded by the David and Lucile Packard Foundation, the Heising-Simons Foundation, the Gemini Observatory, and the University of Chicago. We thank the staff of the Gemini Observatory for their assistance with the commissioning and operation of the instrument. J.L.B.\ and M.R.L.\ acknowledge support from NASA XRP grant 80NSSC19K0293. This work was enabled by observations made from the Gemini North telescope, located within the Maunakea Science Reserve and adjacent to the summit of Maunakea. We are grateful for the privilege of observing the Universe from a place that is unique in both its astronomical quality and its cultural significance.

\facilities{Gemini-North (MAROON-X)}
\software{\texttt{astropy} \citep{astropy:2018}, \texttt{barycorrpy} \citep{barycorrpy}, \texttt{matplotlib} \citep{matplotlib:2007}, \texttt{numpy} \citep{numpy:2020}, \texttt{python} \citep{python3:2009}, \texttt{scipy} \citep{2020SciPy-NMeth}}
\newline

\bibliography{kelt9_dayside_2021}{}
\bibliographystyle{aasjournal}

\end{document}